**Title:** Selective generation of reactive oxygen species in plasma activated water using $CO_2$ plasma

**Authors**

Vikas Rathore[1,2*] and Sudhir Kumar Nema[1,2]

1. Atmospheric Plasma Division, Institute for Plasma Research (IPR), Gandhinagar, Gujarat 382428, India

2. Homi Bhabha National Institute, Training School Complex, Anushaktinagar, Mumbai 400094, India

*Email: vikas.rathore@ipr.res.in

**Abstract**

In the present work, a process of a selective generation of reactive oxygen species (ROS) such as $H_2O_2$ and dissolved $O_3$ in plasma-activated water (PAW) is discussed. For selective ROS generation, pure $CO_2$ was used as a plasma-forming gas. The gases species present in plasma and properties of PAW are compared in details when $CO_2$ and air are used as plasma forming gas. The results reveal that PAW ($CO_2$) has significantly higher pH, and low oxidizing potential and electrical conductivity compared to PAW (air). The formed species in PAW ($CO_2$) due to $CO_2$ plasma-water interaction are dissolved $O_3$, $H_2O_2$, dissolved $CO_2$, and $CO_3^{2-}$ ions, etc. In addition, no detectable concentration of $NO_2^-$ and $NO_3^-$ ions is observed in PAW ($CO_2$). PAW ($CO_2$) has a substantially higher concentration of $H_2O_2$ compared to PAW (air). Moreover, increasing plasma treatment time with water significantly increases $H_2O_2$ and dissolved $O_3$ concentration in PAW ($CO_2$). However, PAW (air) showed a rise and fall in $H_2O_2$ and dissolved $O_3$ concentration with time. In conclusion, selective generation of ROS in PAW is possible using $CO_2$ as plasma-forming gas.

**Keywords:** Plasma activated water, $CO_2$ plasma, $CO_2$ emission spectra, reactive oxygen-nitrogen species,

**Introduction**

The plasma-activated water (PAW) technology is one of the fastest growing novel technologies in cold plasma field. This is due to its continuously evolving applications in the field of plasma medicine, plasma agriculture, plasma food science and technology, etc. (1-11). These applications of PAW are possible due to the presence of various stable reactive oxygen-nitrogen species (RONS) in it. These RONS in PAW are formed as a resultant product of plasma-water interaction. Moreover, it brings a physicochemical change in PAW properties such as pH, oxidizing potential, and electrical conductivity, etc. (4, 9-16)

Different RONS have different significance in their applications (2, 4, 5, 17). Such as species like $H_2O_2$, dissolved $O_3$, HO·, and $ONOO^-$ have applications in microbial inactivation, selective killing of cancer cells, enhancing the shelf life of various food products such as fruits, vegetables, meat, seafood, and dairy products, etc. (1, 3-5, 8-11, 17, 18) Moreover, reactive nitrogen species can be used as a nitrogen replacement source for numerous agriculture applications (2, 7, 15, 19-21).

The PAW produced due to plasma-water exposure contains both the reactive oxygen species ($H_2O_2$, dissolved $O_3$, HO·, etc.) and reactive nitrogen species ($NO_2^-$ ions, $NO_3^-$ ions, etc.) (4, 13, 22). This is due to conventionally used plasma forming gases during PAW production which contain oxygen and nitrogen molecules or ionization of surrounding air by noble gas. The frequently used plasma-forming gases at atmospheric pressure during PAW production are air, nitrogen ($N_2$), oxygen ($O_2$), argon (Ar), Helium (He), and their mixture in different compositions (4, 7, 9, 11, 13, 18, 23). Working with 100% $O_2$ plasma at atmospheric pressure is not recommended. Since, $O_2$ is highly oxidizing and can ignite flammable material rapidly that can cause explosions at atmospheric pressure. The use of inert gases (Ar or He) for plasma production mainly discharges the atmospheric air resulting in the generation of various RONS during plasma-water exposure (23). Hence, the production of plasma-activated water with a selective generation of ROS (free from nitrogen species) at atmospheric pressure is still an open challenge to be overcome. The presence of RNS forms nitrous and nitric acid (strong acid) in PAW due to which the pH of PAW decreased substantially. Therefore, PAW cannot be used in applications that do not prefer low pH solutions. Due to this applicability of PAW is restricted and also one of the main disadvantages of PAW. As per the authors' knowledge, no work has been reported that emphasizes the selective generation of reactive oxygen species (ROS). The ROS have applications in microbial and biofilm inactivation, medicine, food preservation, and enhancing seeds germination, etc (2-4, 8, 9, 24). This research gap tries to be fulfilled in present work.

The conventionally used oxidizing and inert gases plasma formed nitrogen species in water as discussed above (4, 7, 9, 11, 13, 18, 23). Moreover, nitrogen free gases such as phosphine ($PH_3$), hydrogen sulphide ($H_2S$), arsine ($AsH_3$) and sulphur dioxide ($SO_2$), etc. have environmental hazards (highly toxic) at atmospheric pressure, hence not recommended. Therefore, the present work uses $CO_2$ as plasma forming gas for selective generation of reactive oxygen species in plasma-activated water. Moreover, the PAW produced using $CO_2$ plasma is also compared with PAW produced using air plasma. The comparison was performed based on plasma characterization, formation of plasma reactive species/radicals, and properties of PAW.

**Material and Methods**

**Experimental Setup**

The experimental schematic of characterization of air and $CO_2$ plasma and production of plasma-activated water (PAW) is shown in figure 1. The air and $CO_2$ plasma were produced in a co-axial cylindrical pencil plasma jet (PPJ)(25). The PPJ setup based on the principle of dielectric barrier discharge and the schematic is shown in figure 1. In which the central ground electrode was made using a 1.6 mm tungsten rod. The high voltage electrode was made from copper in cylindrical form with an inner diameter of 6 mm in which dielectric quartz tube (outer diameter × inner diameter – 6 mm × 4 mm) was tightly fixed. The discharge gap between the dielectric surface and the ground electrode was 1.2 mm.

To measure the voltage drop across the PPJ setup a 1000x voltage probe (Tektronix P6015A) and a 100 MHz bandwidth, 2 Gs s$^{-1}$ sampling rate, and a 4-CH oscilloscope (Tektronix TDS2014C) was used (26). A 10x (Tektronix TPP0201) voltage probe was used to measure the current and transported charge. This probe measured the voltage drop across the resistor (R - 30 Ω) and capacitor (C - 100 µF) connected in series with the ground as shown in figure 1. The air and $CO_2$ plasma emission spectra were measured by capturing the afterglow light photons using optical fiber and a spectrometer (Plasma and Vacuum Solution (PVS), model UVH-1) as shown in figure 1.

For plasma-activated water production, 50 ml of ultrapure milli-Q water was taken in 600 ml of a glass beaker. This water was treated with air and $CO_2$ plasma as shown in figure 1. The air and $CO_2$ gas flow rate were controlled using a flow controller and the flow rate was fixed at 3 l min$^{-1}$. To enhance the solubility of plasma produced reactive species in water and escalate the reaction between gases reactive species and water, a continuous stirring of water and cooling of water were performed. For water stirring, a mortarless magnetic stirrer was used and for cooling of water, ice-cooled water was placed in contact with a glass beaker in which PAW was kept during plasma-water interaction.

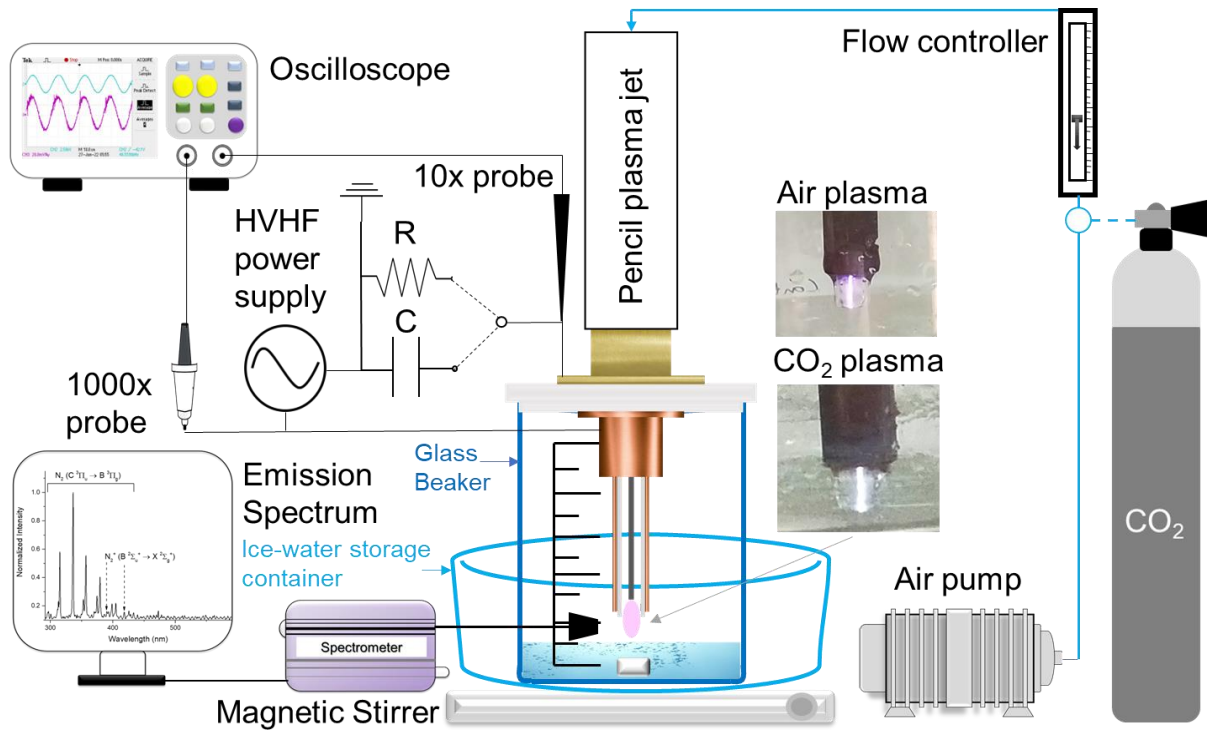

**Figure 1.** Schematic of production of plasma activated water using $CO_2$ and air plasma and these gases plasma electrical and optical emission characterization.

**Equipments used for measurement of physicochemical properties of PAW**

The physicochemical properties of PAW (air or $CO_2$) such as pH, oxidation-reduction potential (ORP), total dissolved solids (TDS), and electrical conductivity (EC) were measured for PAW characterization. A Hanna Instruments pH meter (HI98121), HM digital ORP meter (ORP-200), HM digital TDS meter (AP-1), and Contech Instruments Ltd. EC meter (CC-01) were used to measure the pH, ORP, TDS, and EC of PAW.

**Measurement of Reactive Oxygen-Nitrogen Species Concentration**

The reactive oxygen-nitrogen species (RONS) form in PAW (air or $CO_2$) due to air plasma or $CO_2$ plasma water interaction was determined semi-quantitatively and quantitatively. Strip test and colorimetry test kits were used to determine the initial RONS concentration in PAW and plasma (VISOCOLOR alpha (MACHEREY-NAGEL item no. 935065) nitrate ($NO_3^-$) ions colorimetry test kit, dissolved $O_3$ test kit (Hanna Instruments item no. HI38054), $H_2O_2$ determination test strips (QUANTOFIX Peroxide 25, MACHEREY-NAGEL item no. 91319), $NO_2^-$ ions determination test strips (QUANTOFIX Nitrite, MACHEREY-NAGEL item no. 91311), gases $O_3$ determination test strips (Ozone Test for Ozone in air, MACHEREY-NAGEL item no. 90736)).

The quantitative estimation of RONS concentration present in PAW was measured spectrophotometrically. The $NO_3^-$ ions concentration was measured using the ultraviolet screening method (27), and the standard curve of $NO_3^-$ ions was made using $NaNO_3$ solution with molar attenuation coefficient 0.0602 (mg $L^{-1}$)$^{-1}$(25). In acidic region, $NO_2^-$ ions present in the solution when react with the reaction mixture of N-(1-naphthyl) ethylenediamine dihydrochloride and sulfanilamide give reddish purple azo dye ($\lambda_{max}$ = 540 nm)(27). This characterstic of $NO_2^-$ ions was utilize to determine its unknown concentration. The standard curve of $NO_2^-$ ions was made using $NaNO_2$ solution with a molar attenuation coefficient of 0.0009 (µg $L^{-1}$)$^{-1}$(25). The unknown concentration of $H_2O_2$ in PAW was determined spectrophotometrically using the titanium sulfate method (13). The standard curve of $H_2O_2$ was made of 30% $H_2O_2$ solution (molar attenuation coefficient 0.4857 $mM^{-1}$(25)). The unknown concentration of dissolved $O_3$ in PAW was determined using the indigo colorimetric method (27).

The titratable acidity and dissolved $CO_2$ concentration in PAW ($CO_2$) were determined using the titration method(8, 28). The titratable acidity of PAW ($CO_2$) was determined using 0.1 M sodium hydroxide (NaOH) solution and a freshly prepared phenolphthalein indicator. In addition, the dissolved $CO_2$ concentration in PAW was determined using 0.02 N sodium carbonate ($Na_2CO_3$) solution and freshly prepared phenolphthalein as an indicator (28). The dissolved carbonate ions ($CO_3^{2-}$) concentration in PAW ($CO_2$) was determined using the UV screening method (29, 30) with $CO_3^{2-}$ molar attenuation coefficient 0.0008 (mg $L^{-1}$)$^{-1}$.

**Data analysis**

All the experimental were performed at least three times (n ≥ 3) in the present investigation. The results were shown as μ ± σ (mean ± standard deviation (Error)). The statistically significant difference with a significant level of 95% (p = 0.05) among the groups mean were calculated using one-way analysis of variance followed by a post-hoc test (Fischer Least Significant Difference (LSD)).

**Results and discussion**

**Voltage current waveform**

The air and $CO_2$ plasma voltage current waveform produced in dielectric barrier discharge (DBD) pencil plasma jet (PPJ) is shown in figure 2. The air and $CO_2$ plasma current waveform showed nanosecond (ns) current filaments peaks (~ 100 ns, shown in figure S1 of supplementary material) over each negative and positive current half cycle. The cluster of these nanosecond current filaments lies in the microsecond region. Therefore, these current discharges are known as filamentary DBD micro discharges (26). The discharge current peaks observed in $CO_2$ plasma were higher than air plasma for the same applied voltage. This signifies the radicals and species produced in $CO_2$ plasma had higher current carrying affinity compared to air plasma. Alongside, the other possible region is the generation of high concentration of plasma radicals and species in $CO_2$ plasma compared to air plasma. Due to which high discharge current was observed in $CO_2$ plasma for the same process parameter.

The plasma discharge power consumed during the air and $CO_2$ plasma generation was measured using voltage-transported charge Lissajous figure. Initially, the energy consumed during discharge was calculated using the integral of voltage over the charge domain. For power calculation, a product of energy consumed with frequency (40 kHz) was performed. To compare the properties of PAW produced using air and $CO_2$ plasma, the energy and power kept in the range of 0.0125 mJ to 0.015 mJ and 0.5 to 0.6 W.

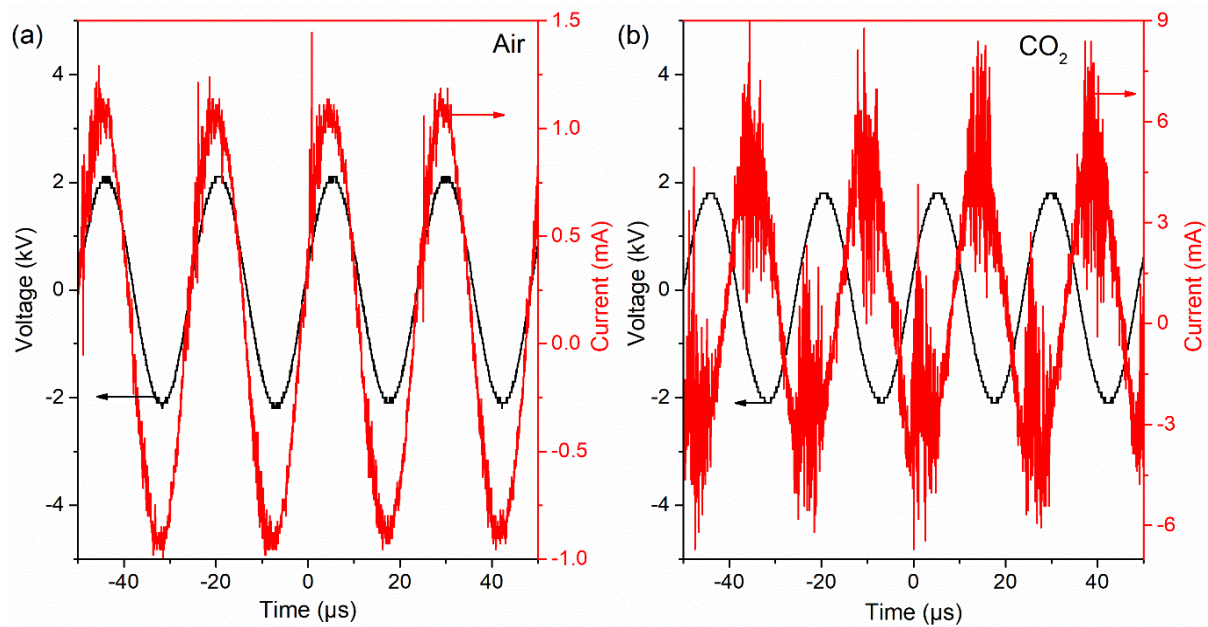

**Figure 2.** Voltage current waveform of (a) air and (b) $CO_2$ plasma produced in pencil plasma jet

**Optical emission spectra of air and $CO_2$ plasma**

The optical emission spectra of air and $CO_2$ plasma in the afterglow region are shown in figure 3. The overlay plot of air (solid line) and $CO_2$ (dotted lines) plasma showed the deviation between the emission bands peaks lines of air and $CO_2$ plasma. These emission bands peaks lines are formed due to electronic transition radiative decay of the upper vibration state to the lower vibration state of ions or molecules. The emission spectrum of air plasma consists of strong emission band peaks of $N_2$ second positive system (C $^3\Pi_u \rightarrow$ B $^3\Pi_g$) along with weak emission intensity band peaks of $N_2^+$ first negative system (B $^2\Sigma_u^+ \rightarrow$ X $^2\Sigma_g^+$)(31, 32). The reactions associated with this transition are shown in equations (1-4). Moreover, the $CO_2$ plasma afterglow region showed strong intensity emission band peaks of $CO_2^+$ first negative system (A $^2\Pi_g \rightarrow$ X $^2\Pi_u$). Along with that weak intensity emission band peaks of $CO^+$ (A $^2\Pi$

→ X $^2\Sigma$), CH (A $^2\Delta$ → X $^2\Pi$), CO (d $^3\Delta$ → a $^3\Pi$), and C$_2$ (A $^3\Pi_g$ → X $^3\Pi_u$) also observed in CO$_2$ plasma afterglow region (33). The formation of these upper state ions and molecules in the air and CO$_2$ plasma region occurs due to the collision of gas (air or CO$_2$) with high-energy electrons, photons, ions, and neutral particles. These collisions result in the excited/upper vibration state of the above molecules. The most common reactions involved in these collisions were excitation, ionization, and dissociation (33-36). The details of air and CO$_2$ emission band peaks lines observed in the air and CO$_2$ plasma afterglow region are shown in Table S1 of supplementary material.

For air emission spectrum:

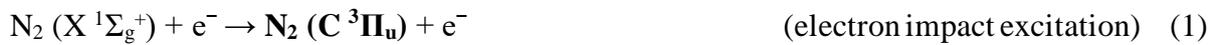
$$N_2\ (X\ ^1\Sigma_g^+) + e^- \rightarrow N_2\ (C\ ^3\Pi_u) + e^- \qquad \text{(electron impact excitation)} \quad (1)$$

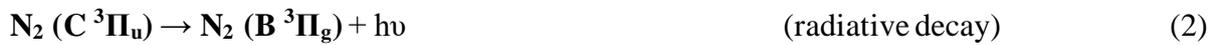
$$N_2\ (C\ ^3\Pi_u) \rightarrow N_2\ (B\ ^3\Pi_g) + h\upsilon \qquad \text{(radiative decay)} \quad (2)$$

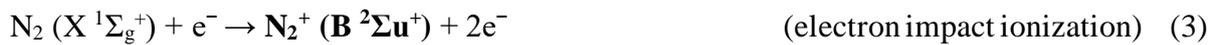
$$N_2\ (X\ ^1\Sigma_g^+) + e^- \rightarrow N_2^+\ (B\ ^2\Sigma u^+) + 2e^- \qquad \text{(electron impact ionization)} \quad (3)$$

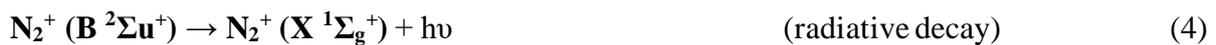
$$N_2^+\ (B\ ^2\Sigma u^+) \rightarrow N_2^+\ (X\ ^1\Sigma_g^+) + h\upsilon \qquad \text{(radiative decay)} \quad (4)$$

For CO$_2$ emission spectrum:

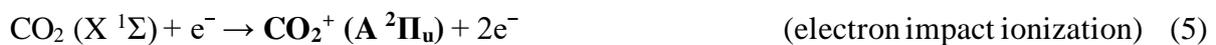
$$CO_2\ (X\ ^1\Sigma) + e^- \rightarrow CO_2^+\ (A\ ^2\Pi_u) + 2e^- \qquad \text{(electron impact ionization)} \quad (5)$$

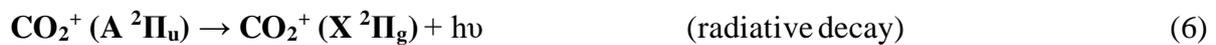
$$CO_2^+\ (A\ ^2\Pi_u) \rightarrow CO_2^+\ (X\ ^2\Pi_g) + h\upsilon \qquad \text{(radiative decay)} \quad (6)$$

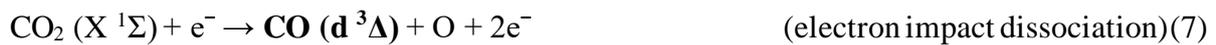
$$CO_2\ (X\ ^1\Sigma) + e^- \rightarrow CO\ (d\ ^3\Delta) + O + 2e^- \qquad \text{(electron impact dissociation)}(7)$$

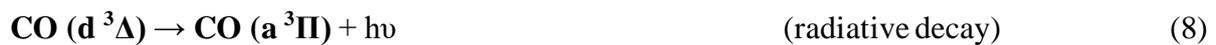
$$CO\ (d\ ^3\Delta) \rightarrow CO\ (a\ ^3\Pi) + h\upsilon \qquad \text{(radiative decay)} \quad (8)$$

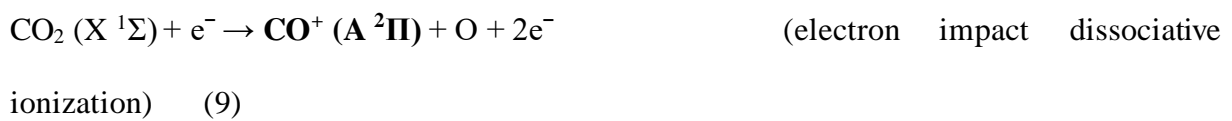
$$CO_2\ (X\ ^1\Sigma) + e^- \rightarrow CO^+\ (A\ ^2\Pi) + O + 2e^- \qquad \text{(electron impact dissociative ionization)} \quad (9)$$

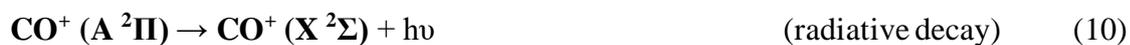
$$CO^+\ (A\ ^2\Pi) \rightarrow CO^+\ (X\ ^2\Sigma) + h\upsilon \qquad \text{(radiative decay)} \quad (10)$$

$CO_2 (X\ ^1\Sigma) + e^- \rightarrow C + O_2 + e^-$   (electron impact dissociative ionization)   (11)

$C + C \rightarrow \mathbf{C_2\ (A\ ^3\Pi_g)}$   (Recombination)   (12)

$\mathbf{C_2\ (A\ ^3\Pi_g) \rightarrow C_2\ (X\ ^3\Pi_u)}$   (radiative decay)   (13)

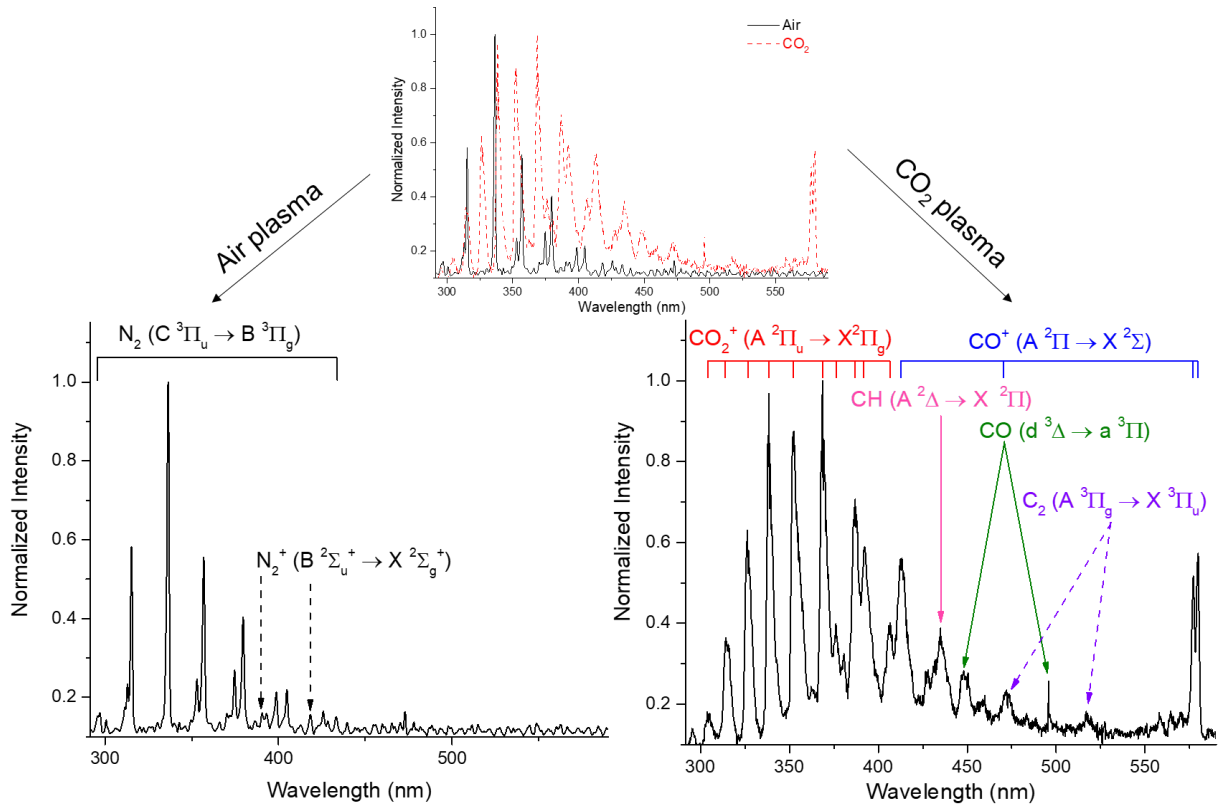

**Figure 3.** Air and $CO_2$ plasma emission spectra recorded in afterglow region

**Physicochemical properties of plasma-activated water**

The plasma-activated water (PAW) produced using air and $CO_2$ plasma is colorless in appearance. However, PAW ($CO_2$) and PAW (air) can be differentiated based on odor. PAW (air) was odorless, but PAW ($CO_2$) has a smoky and unpleasant odor.

The physicochemical properties of PAW produced using air and $CO_2$ plasma is shown in figure 4. The pH of PAW produced using air and $CO_2$ plasma showed a continuous decrease in its value with increasing plasma treatment time (figure 4 (a)). This signifies increasing plasma-water treatment time results in more formation of acidic radicals/species in water (17, 22, 25). Moreover, the PAW produced using $CO_2$ plasma was considerably higher (82.14% higher after 60 minutes of treatment) than PAW produced using air plasma. Hence, the generated species in water when exposed to air plasma were more acidic compared to $CO_2$ plasma. Ma et al. (9) Subramanian et al. (1), El Shaer et al. (11), and Lu et al. (16) also showed a decrease in pH of PAW with increasing plasma treatment with water.

The oxidizing potential of PAW when produced using air and $CO_2$ plasma is shown in figure 4 (b). The oxidizing potential of PAW gives information regarding net combinations of oxidizing and reducing species formed in water after plasma exposure (12, 23, 25). The oxidation-reduction potential (ORP) of PAW prepared using air plasma was higher compared to PAW prepared using $CO_2$ plasma. This was due to the formation of a high concentration of oxidizing species in PAW when produced using air plasma compared to $CO_2$ plasma. For 60 minutes of plasma treatment, the ORP of PAW prepared using air plasma was 27.1% higher compared to $CO_2$ plasma. The increase in ORP of PAW with increasing plasma-water treatment is also shown in the work reported by Guo et al. (3), Xiang et al. (10), and Ma et al. (9), etc.

The rough estimation of conducting ions were measured by measuring total dissolved solids (TDS) and electrical conductivity (EC). The TDS and EC give the information regarding conducting ions formed in water due to plasma-water interaction. The observed TDS and EC of PAW, when prepared using air and $CO_2$ plasma are shown in figure 4 (c, d). Increasing plasma treatment with water increased the TDS and EC of PAW for both the air and $CO_2$ plasma. The observed TDS and EC of PAW prepared using air plasma were substantially high compared to $CO_2$ plasma (937.0% and 987.3% higher after 60 minutes of treatment). Hence, the concentration of inorganic ions formed in water after air plasma exposure was extremely higher compared to $CO_2$ plasma. The increase in EC with plasma treatment was also supported by results of Subramanian et al. (1), Zhang et al. (37), and Sivachandiran et al. (15), and Lu et al. (16), etc.

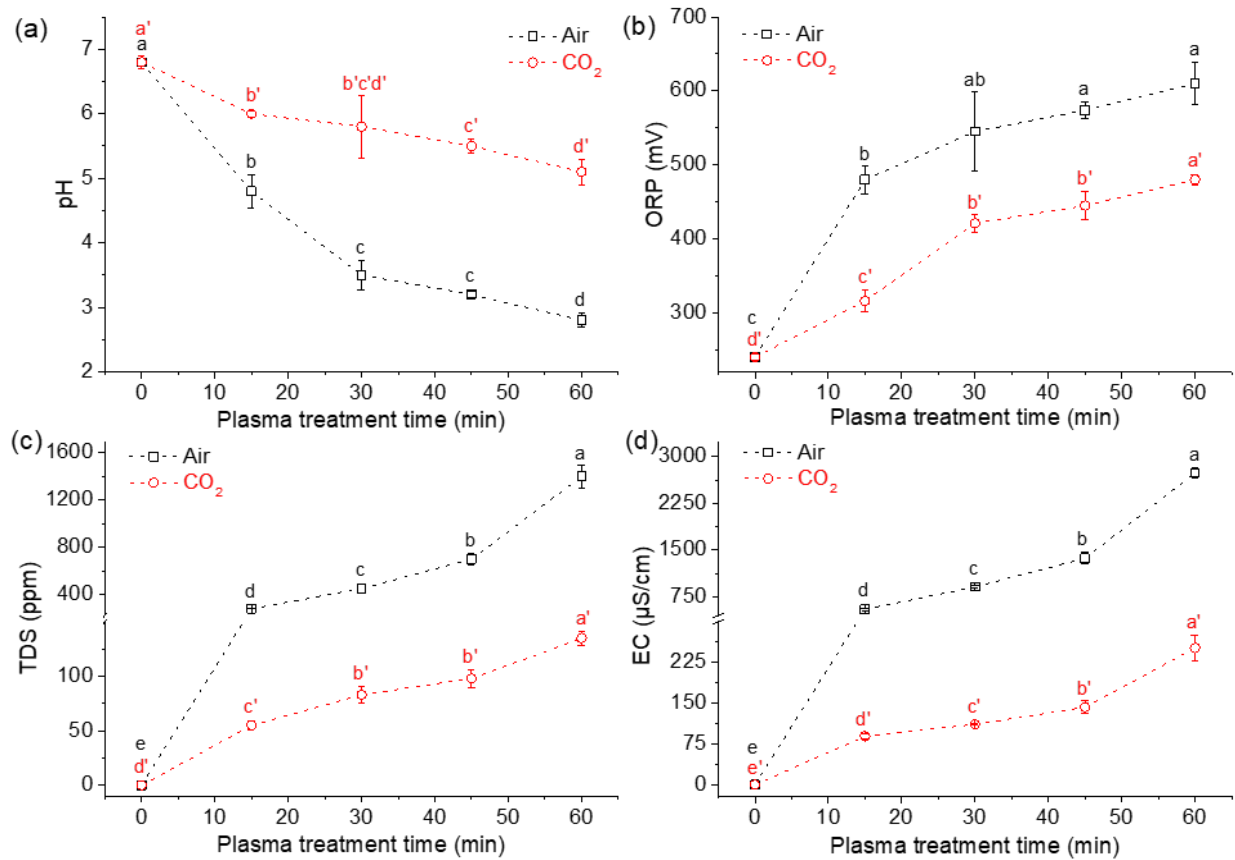

**Figure 4.** The variation in physicochemical properties of plasma activated water prepared using air and $CO_2$ plasma with plasma treatment time. (a) pH, (b) oxidation-reduction potential (ORP), (c) total dissolved solids (TDS), and (d) electrical conductivity (EC). Statistically significant ($p < 0.05$) difference between the group mean ± standard deviation ($\mu \pm \sigma$) is shown by a different lowercase letter.

**RONS concentration in plasma-activated water**

The above-discussed variation in physicochemical properties of water after plasma treatment occurs due to the formation of numerous reactive species in water (1, 3, 4, 6-8, 14, 20, 22). The mechanism of formation of these reactive species in PAW is shown in equations (14-25) (1, 4, 12, 13, 16, 18, 22, 25).

Formation of reactive oxygen species (ROS) in plasma-activated water:

$$O_2(g) \rightarrow 2O(g) \tag{14}$$

$$O_2(g) + O(g) \rightarrow O_3(g) \xrightarrow{aq.} \mathbf{O_3(aq.)} \tag{15}$$

$$H_2O(g) \rightarrow H^\cdot(g) + HO^\cdot(g) \xrightarrow{aq.} \mathbf{H^+(aq.) + e^-(aq.) + HO^\cdot(aq.)} \tag{16}$$

$$2HO^\cdot(aq.) \rightarrow \mathbf{H_2O_2(aq.)} \tag{17}$$

$$\mathbf{H_2O_2(aq.) + O_3(aq.)} \rightarrow HO^\cdot(aq.) + HO_2^\cdot(aq.) + O_2(aq.) \tag{18}$$

$$2HO_2^\cdot(aq.) \rightarrow \mathbf{H_2O_2(aq.)} + O_2(aq.) \tag{19}$$

Formation of reactive nitrogen species (RNS) in plasma-activated water:

$$N_2(g) \rightarrow 2N(g) \tag{20}$$

$$N(g) + xO(g) \rightarrow NO_x(g) \xrightarrow{aq.} NO_x(aq.) \; \{NO(aq.), NO_2(aq.), \; NO_3(aq.), etc.\} \tag{21}$$

$$2NO_2(aq.) + H_2O(aq.) \rightarrow \mathbf{NO_2^-(aq.) + NO_3^-(aq.) + 2H^+(aq.)} \; \{HNO_2(aq.) + HNO_3(aq.)\} \tag{22}$$

$$NO(aq.) + NO_2(aq.) + H_2O(aq.) \rightarrow \mathbf{2NO_2^-(aq.) + 2H^+(aq.)} \; \{HNO_2(aq.)\} \tag{23}$$

$$\mathbf{NO_2^-(aq.) + O_3(aq.)} \rightarrow \mathbf{NO_3^-(aq.)} + O_2(aq.) \tag{24}$$

$$\mathbf{NO_2^-(aq.) + H_2O_2(aq.)} \rightarrow \mathbf{NO_3^-(aq.)} + H_2O(aq.) \tag{25}$$

Figure 5 showed the identified and measured concentration of RONS (reactive oxygen-nitrogen species) present in PAW when prepared using air and $CO_2$ plasma. Figure 5 (a, c, e, g) showed the RONS such as $NO_2^-$ ions, $NO_3^-$ ions, $H_2O_2$, and dissolved $O_3$ in PAW when prepared using air plasma. The reactions involved in the formation of RONS in PAW (air) are given in equations (14-25 (16)). Moreover, the reactive species formed in PAW when using $CO_2$ plasma are given as $H_2O_2$, dissolved $O_3$, dissolved $CO_2$, and $CO_3^{2-}$ ions (equations (14-19, 26-27)) (figure 5 (f, h) and figure 6).

The concentration of $NO_3^-$ and $NO_2^-$ ions present in PAW prepared using air and $CO_2$ plasma is shown in figure 5 (a-d). A continuous increase in $NO_3^-$ and $NO_2^-$ ions concentration with plasma treatment time observed in PAW prepared using air plasma (figure 5 (a, c)). The observed maximum concentration of $NO_3^-$ and $NO_2^-$ ions in PAW (air) were given as 4.0 mg $L^{-1}$ and 401.5 mg $L^{-1}$, respectively. The $NO_3^-$ and $NO_2^-$ ions form nitric and nitrous acid in PAW (air) (equations (20-25)) (1, 6, 8, 12, 22). As nitric acid is a strong acid, therefore the lowest pH value of PAW (air) was given as 2.8. The increasing concentration of $NO_2^-$ and $NO_3^-$ ions with activation time was also reported by Subramanian et al. (1) and Xiang et al. (10) in PAW prepared in an air atmosphere. However, the PAW ($CO_2$) did not contain any observable concentration of $NO_3^-$ and $NO_2^-$ ions as shown in figure 5 (b, d). As discussed in equations (1-4, 20-23), the formation of RNS (reactive nitrogen species) in PAW required excited nitrogen species (12, 18, 22, 25) that were not observed in emission spectra of $CO_2$ plasma (figure 3). Hence, the possible RNS present in PAW ($CO_2$) such as ($NO_2^-$ and $NO_3^-$ ions) were beyond the detection limit of the present investigation.

Moreover, the concentration of $H_2O_2$ present in PAW ($CO_2$) was 316.7% higher than PAW (air). This was due to no interference of $NO_2^-$ ions in $H_2O_2$ determination in PAW ($CO_2$). The $NO_2^-$ ions present in PAW (air) react with $H_2O_2$ to give more stable $NO_3^-$ ions (equation 25) (6, 18, 25). Therefore, interfere with the $H_2O_2$ determination in PAW (air). The interference of $NO_2^-$ ions in $H_2O_2$ concentration and variation can be seen in figure 5 (e). Initially (t = 0 minutes), no $H_2O_2$ was present in PAW (air), as the plasma treatment increased to 30 minutes, a continuous increase in the $H_2O_2$ concentration was observed. Increasing plasma treatment time to 45 minutes results in a decrease in $H_2O_2$ concentration due to the reaction of $NO_2^-$ ions with $H_2O_2$ to give more stable $NO_3^-$ ions. Further increasing plasma treatment time to 60 minutes results in $H_2O_2$ concentration enhancement. This showed saturation of $NO_2^-$ ions and $H_2O_2$ reaction in PAW (air) and the unreacted $H_2O_2$ shown by enhanced $H_2O_2$ in PAW (air). Similar behavior as $H_2O_2$ was observed in the concentration of dissolved $O_3$ in PAW (air). Since, $NO_2^-$ ions present in PAW (air) also reacts with dissolved $O_3$ to give more stable $NO_3^-$ ions by following equation (24). This rise and fall in $H_2O_2$ concentration in PAW (air) with increasing plasma treatment time also was observed in work reported by Subramanian et al. (1) and Sivachandiran et al. (15).

However, this rise and fall in $H_2O_2$ and dissolved $O_3$ concentration in PAW prepared using $CO_2$ plasma was not observed due to the absence of $NO_2^-$ ions (figure 5 (b, d, g, h)). As no $N_2$ emission peaks bands were observed in the $CO_2$ plasma (figure 3) that confirms the absence of nitrogen species in $CO_2$ plasma. Hence, no interference of $NO_2^-$ ions in PAW ($CO_2$) results in a monotonous increase in $H_2O_2$ and dissolved $O_3$ concentration with increasing plasma treatment time with water (figure 5 (g, h)).

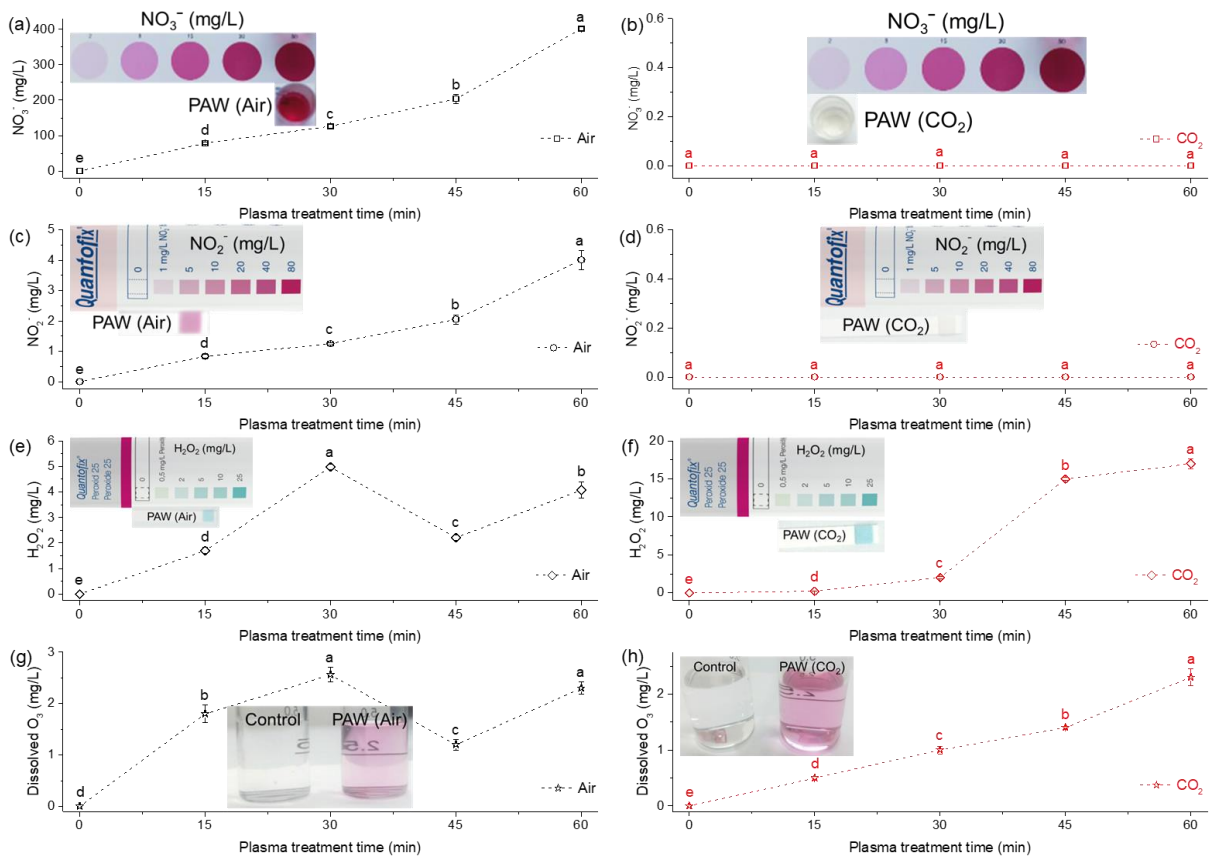

**Figure 5.** The variation in reactive oxygen-nitrogen species (RONS) concentration of plasma-activated water prepared using air and $CO_2$ plasma with plasma treatment time. (a, b) $NO_3^-$ ions, (c, d) $NO_2^-$ ions, (e, f) $H_2O_2$ concentration, and (g, h) Dissolved $O_3$. Statistically significant ($p < 0.05$) difference between the group mean ± standard deviation ($\mu \pm \sigma$) is shown by a different lowercase letter.

The variation in titratable acidity, dissolved $CO_2$, and $CO_3^{2-}$ ions with plasma treatment time in PAW ($CO_2$) is shown in figure 6. The excited carbon oxides ($CO_x$) and carbon oxide ions ($CO_x^+$), etc. observed in emission spectra of $CO_2$ plasma (figure 3) when comes in contact with water enhances the solubility of $CO_2$ and formed carbonic acid ($H_2CO_3$), etc. in water (equations (26-27)(38)). Due to which physicochemical properties of PAW ($CO_2$) changed. The acidic species concentration formed in PAW ($CO_2$) was measured by measuring titratable acidity. Increasing plasma treatment time with water continuously and significantly ($p < 0.05$) increases the titratable acidity, dissolved $CO_2$, and $CO_3^{2-}$ ions concentration. The uniform increase in titratable acidity, dissolved $CO_2$, and $CO_3^{2-}$ ions concentration signifies continuous production of reactive species in PAW ($CO_2$) with increasing plasma-water treatment time. The $CO_3^{2-}$ ions exist in the form of carbonic acid in PAW ($CO_2$). The dissolved $CO_2$ and $CO_3^{2-}$ ions (carbonic acid) are weak acids due to which the pH of PAW ($CO_2$) decreased. However, this decreases in pH of PAW ($CO_2$) significantly ($p < 0.05$) low compared to PAW (air).

Formation of carbonic acid in PAW:

$$H\ (aq.) + CO_2(aq.) \rightarrow HOCO(aq.) \tag{26}$$

$$HOCO\ (aq.) + OH(aq.) \rightarrow \boldsymbol{H_2CO_3(aq.)} \tag{27}$$

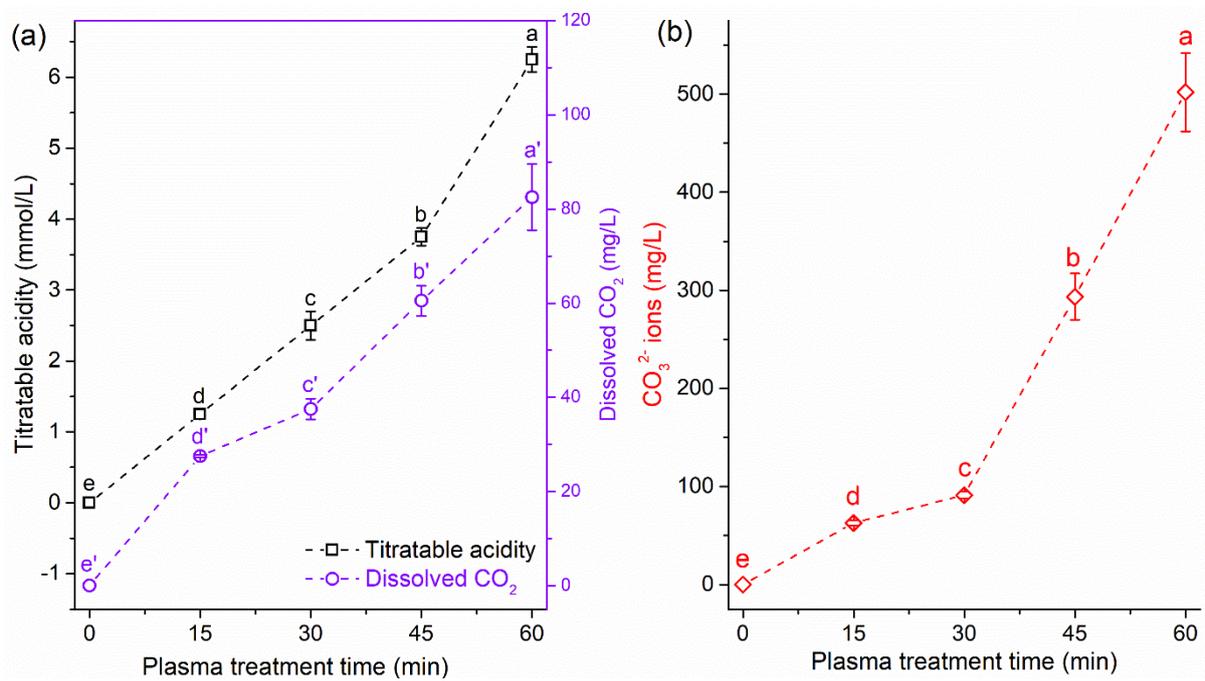

Figure 6. (a) Titratable acidity and dissolved $CO_2$, and (b) $CO_3^{2-}$ ions concentration in plasma-activated water produced using $CO_2$ plasma. Statistically significant ($p < 0.05$) difference between the group mean ± standard deviation ($\mu \pm \sigma$) is shown by a different lowercase letter.

Hence, the above results and discussion showed the higher discharge current filaments in $CO_2$ plasma compared to air plasma. Moreover, the emission spectrum of $CO_2$ plasma is free from nitrogen containing species. As a results, formation of reactive nitrogen species (RNS) is not occurring in PAW ($CO_2$). Hence, selective generation of reactive oxygen species (ROS) occurs in PAW ($CO_2$). Moreover, due to the use of $CO_2$ gas plasma for PAW preparation. The carbonic acid, dissolved $CO_2$, $CO_3^{2-}$ ions also occurs in PAW due to which pH of PAW ($CO_2$) is decreased. However, the pH of PAW ($CO_2$) is significantly lower than PAW (air).

**Conclusion**

The present work compares the properties of PAW produced using air and $CO_2$ plasma. The acidity of PAW (air) is significantly higher than PAW ($CO_2$). This is due to the dissolution of strong acids (nitric acid) in PAW (air) compared to weak acids (carbonic acid) of PAW ($CO_2$). In addition, the oxidizing potential, total dissolved solids, and electrical conductivity of PAW (air) are significantly higher than PAW ($CO_2$). This is due to PAW (air) has high concentration of strong ionic species in the form of $HNO_3$ compared to weak $H_2CO_3$ species of PAW ($CO_2$). The PAW prepared using $CO_2$ plasma does not contain any reactive nitrogen species. This is due to the emission spectra of $CO_2$ plasma not containing any $N_2$ emission band peaks. Hence, $CO_2$ plasma-water interaction does not form any reactive nitrogen species in PAW ($CO_2$). Hence, selective production of reactive oxygen species can be achieved without the interference of reactive nitrogen species. Therefore, the concentration of dissolved $H_2O_2$ in PAW ($CO_2$) is higher than PAW (air). In conclusion, selective production of reactive oxygen species in plasma-activated water is possible by using $CO_2$ as a plasma-forming gas. The presence of reactive oxygen species in PAW ($CO_2$) makes it a useful antimicrobial agent. Moreover, it can also be used in numerous applications where conventional PAW could not be used due to its low pH (such as low pH PAW could not be used for surface disinfection of metal objects since it oxidizes its surface and damage it).

**Acknowledgments**

This work was supported by the Department of Atomic Energy (Government of India) doctrate fellowship scheme (DDFS).

**Data availability statement**

The data that support the findings of this study are available upon reasonable request from the authors.

**Conflict of interests**

The authors declare that there are no conflicts of interests.

**Authors' contributions**

Both authors contributed to the study conception and design. Material preparation, data collection, and analysis were performed by Vikas Rathore. The first draft of the manuscript was written by Vikas Rathore, and both authors commented on previous versions of the manuscript. Both authors read and approved the final manuscript.

**ORCID iDs**

Vikas Rathore https://orcid.org/0000-0001-6480-5009